\documentstyle[11pt,newpasp,twoside,epsf]{article}
\begin{document}
\title{The PLANET microlensing campaign: 
Implications for planets around galactic disk and bulge stars}
\author{M. Dominik\altaffilmark{1}, M. D. Albrow\altaffilmark{2}, J.-P. Beaulieu\altaffilmark{3}, 
J. A. R. Caldwell\altaffilmark{4}, A. Cassan\altaffilmark{3},
C. Coutures\altaffilmark{5}, J. Greenhill\altaffilmark{6}, K. Hill\altaffilmark{6},
P. Fouqu\'{e}\altaffilmark{5},
K. Horne\altaffilmark{1}, U. G. J{\o}rgensen\altaffilmark{7}, S. Kane\altaffilmark{1}, 
D. Kubas\altaffilmark{8}, R. Martin\altaffilmark{9}, J. Menzies\altaffilmark{10},
K. R. Pollard\altaffilmark{2}, 
K. Sahu\altaffilmark{4},
J. Wambsganss\altaffilmark{8}, R. Watson\altaffilmark{6}, A. Williams\altaffilmark{9}}
\affil{\altaffilmark{1}University of St Andrews,
School of Physics \& Astronomy,
North Haugh, St Andrews, KY16 9SS, United Kingdom}
\affil{\altaffilmark{2}University of Canterbury,  Dept. of Physics \& Astronomy, Private Bag 4800,
Christchurch, New Zealand}
\affil{\altaffilmark{3}Institut d'Astrophysique de Paris, 98bis Boulevard Arago,
75014 Paris, France}
\affil{\altaffilmark{4}Space Telescope Science Institute, 3700 San Martin Drive, Baltimore,
MD 21218, U.S.A.}
\affil{\altaffilmark{5}European Southern Observatory, 
Casilla 19001, Santiago 19, Chile}
\affil{\altaffilmark{6}Univ. of Tasmania, Physics Dept., G.P.O. 252C, Hobart, Tasmania 7001,
Australia}
\affil{\altaffilmark{7}Astronomisk Observatorium, Niels Bohr Institutet, K{\o}benhavns
Universitet, Juliane Maries Vej 30, 2100 K{\o}benhavn {\O}, Denmark}
\affil{\altaffilmark{8}Universit\"{a}t Potsdam, Institut f{\"ur} Physik,
Am neuen Palais 10, 14469 Potsdam, Germany}
\affil{\altaffilmark{9}Perth Observatory, Walnut Road, Bickley, Perth 6076, Australia}
\affil{\altaffilmark{10}South African Astronomical Observatory, P.O. Box 9, Observatory 7935,
South Africa}

\begin{abstract}
With round-the-clock monitoring of galactic bulge microlensing events, 
the PLANET experiment
constrains the abundance and can yield the discovery of planets down to the mass of earth 
around galactic disk and bulge stars. 
Data taken until 1999 imply that
less than 1/3 of bulge M-dwarfs are surrounded by jupiter-mass companions at 
orbital radii between 1 and 4~AU.
The current rate of microlensing alerts allows 15--25 jupiters and 1--3
earths to be probed per year.

\end{abstract}

Microlensing is sensitive to unseen planets of mass $m$ at a projected orbital
radius $r_{\rm p}$ around unseen lens stars 
of mass $M$, mainly M-dwarfs ($M \sim 0.3~M_\odot$), at the distance $D_{\rm L}$ 
that cause microlensing events of durations $\sim\,1$~month on background source stars 
at the distance $D_{\rm S}$. Distortions of 1--20\,\% to
the microlensing light curve caused by a planet last 
from hours (earth) to days (jupiter) 
and their probability
increases towards 
$r_{\rm p} \sim r_{\rm E} \sim 2.5~\mbox{AU}$, where
$r_{\rm E} = \sqrt{2\,R_{\rm S}\,D}$ denotes the Einstein radius of the lens star,
with $R_{\rm S} = (2GM)/c^2$ being the Schwarzschild radius and
$D = D_{\rm L}\,(D_{\rm S}-D_{\rm L})/D_{\rm S}$.

With generous time allocations at the observatories, 
PLANET (Probing Lensing
Anomalies NETwork) obtained  
a dense round-the-clock
coverage of galactic bulge microlensing events in $I$ 
with additional observations in $R$ and $V$ with 
its current network of 1m-class telescopes 
formed by 
SAAO 1.0m (South Africa),
Danish 1.54m at ESO LaSilla (Chile),
Canopus 1.0m (Tasmania),
and Perth 0.6m (Western Australia), and also with
Dutch 0.9m and 2.2m at ESO La Silla,
0.9m and Yale 1.0m at CTIO (Chile), and
MSO 50'' (Australia).

The 
photometric precision of 1--2\,\%, dictating the exposure time with the 
target brightness, and a sampling interval 
of 1.5--2.5~hrs, allowing a {\em characterization} 
of distortions by jupiters, 
limit the number of events monitored to up to 20 events at the same time
or 75 events per season (Dominik et al. 2002).

The target monitored 
at any given time is
selected with the aim to maximize the planet detection efficiency.
Events with large peak magnification $A_0$ are preferred 
but spending the whole observing time on such events
is not optimal 
for obtaining results on the abundance of jupiters (Horne 2003), 
whereas all the information about earths will arise 
from events with $A_0 \ga 80$ only. 

Currently, OGLE-III provides $\sim 500$ and MOA provides
$\sim 60$ microlensing alerts per year, which will allow
PLANET to probe 15--25 jupiters and 1--3 earths per year. If no 
planetary distortions are observed, the abundance limits from three
years of observations will be 4--7\,\%  for 
jupiters and $\sim 40\,$\% for earths.

The figure shows the limits on the abundance of planets 
resulting from monitoring 42
events by PLANET between 1995 and 1999 (Gaudi et al. 2002).

\begin{figure}
\plotfiddle{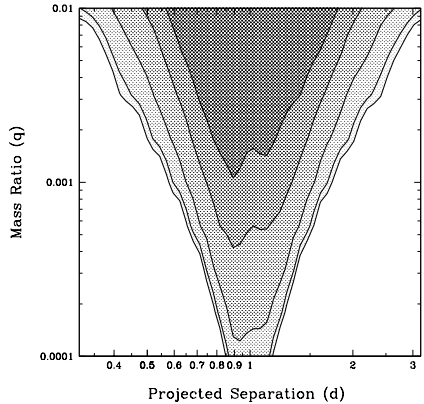}{4.5cm}{0}{30}{30}{-90}{-60}
\caption{
Fractions $f(d,q) =$ ~3/4, 2/3, 1/2, 1/3, and 1/4 (inside to outside)
of lens stars surrounded by a companion with mass ratio
$q = m/M$ at $d = r_{\rm p}/r_{\rm E}$
that are excluded at 95$\,$\% C.L.}
\end{figure}


\begin{references}
\reference Dominik, M., et al. 2002, P\&SS, 50, 299
\reference Gaudi, B. S., et al. 2002, ApJ, 566, 463 
\reference Horne, K. 2003, MNRAS, submitted 
\end{references}
\end{document}